\documentstyle[aps,pre,epsfig]{revtex}

\newcommand{\be}{\begin{equation}}
\newcommand{\ee}{\end{equation}}
\newcommand{\bc}{\begin{center}}
\newcommand{\ec}{\end{center}}
\newcommand{\bi}{\begin{itemize}}
\newcommand{\ei}{\end{itemize}}
\newcommand{\ba}{\begin{eqnarray}}
\newcommand{\ea}{\end{eqnarray}}
\newcommand{\ie}{{\it i.e.\ }}

\newcommand{\ignore}[1]{}

\begin{document}
\draft
\twocolumn[\hsize\textwidth\columnwidth\hsize\csname
@twocolumnfalse\endcsname

\title{Epidemic threshold in structured scale-free networks}

\author{V\'{\i}ctor M. Egu\'{\i}luz$^{1,2,3}$\cite{vme} and
Konstantin Klemm$^{1,3,}$}

\address{
$^1$Instituto Mediterr\'aneo de Estudios Avanzados IMEDEA (CSIC-UIB),
E07071 Palma de Mallorca, Spain
\\$^2$Departamento de F\'{\i}sica, Universidad de las Islas Baleares,
E07071 Palma de Mallorca, Spain
\\$^3$Center for Chaos and Turbulence Studies,
Niels Bohr Institute, Blegdamsvej 17, DK-2100 Copenhagen \O, Denmark
}

\date{\today}

\maketitle

\begin{abstract}

We analyze the spreading of viruses in scale-free networks with high
clustering and degree correlations, as found in the Internet graph. For the
Suscetible-Infected-Susceptible model of epidemics the prevalence undergoes a
phase transition at a finite threshold of the transmission probability.
Comparing with the absence of a finite threshold in networks with purely
random wiring, our result suggests that high clustering and degree
correlations protect scale-free networks against the spreading of viruses.
We introduce and verify a quantitative description of the
epidemic threshold based on the connectivity of the neighborhoods of the
hubs.

\end{abstract}
]


The description of the properties of several real networks has manifested
that, despite their different nature, they share some common features
\cite{Watts98,Barabasi99,Amaral00,Strogatz01}. They typically show a
scale-free distribution of degree, high clustering, and a short average
pathlength \cite{Albert01}. Although their topological properties have been
studied in detail, a natural question that arises is the dynamical properties
that result from the different topologies \cite{Strogatz01}. An example where
the interaction network is crucial for the dynamics is the case of disease
spreading. The study of complex networks as models of social, technological
and biological interaction has been shown to give valuable insights of how
viruses, diseases and rumours spread
\cite{Lloyd01,Pastor01,Kuperman01,Zanette01}.

Most of these investigations have been performed assuming networks with
homogeneous connectivity, where all individuals have approximately the same
number (degree) of contacts with others. The network is typically modeled as
a regular lattice, a random graph, or a superposition of these two
\cite{Watts98}. For such topologies the number of infected individuals
undergoes a phase transition: The single contact transmission probability
needs to exceed a critical threshold for a disease to become epidemic
\cite{Diekmann00,Marro99}.
Recently, however, it has been discovered that many networks involved in the
spread of diseases have a {\rm scale-free} distribution of degree with a
regime of power law decay. In particular, the web of human sexual contacts
\cite{Liljeros01}, the web of electronic mail communication \cite{Ebel02} and
the Internet \cite{Pastor01b} all contain highly connected individuals or
nodes, so-called {\em hubs}, which had been disregarded by the assumption of
homogeneous connectivity in previous works. The first model studies of
disease spread in scale-free networks including hubs have revealed the
absence of an epidemic threshold. Therefore it has been claimed that in
technological and sociological networks even viruses with extremely low
transmission probability can spread, and any prophylactic strategies aiming
at a reduction of the average infectiveness would never result in a total
eradication of a prevalent virus. However, the alarming predictions have been
obtained assuming random mixing.  Apart from the scale-free degree
distribution, all non-trivial topological properties of real-world
sociological and technological networks have been neglected.

This Letter is dedicated to the analysis of virus spreading in networks with local
structure. In order to account for the large clustering coefficient and the
presence of degree correlations \cite{Pastor01b} we model the potentially
infective contacts by highly clustered scale-free networks \cite{Klemm01}.
We find that the single-contact transmission probability needs to exceed
a finite threshold for a virus to spread and prevail. Thus the behavior of
epidemics is qualitatively different in highly clustered scale-free networks
as compared with randomly wired scale-free networks. We conjecture that the
difference can be explained by the presence or absence of connections between
the hubs. Based on this conjecture, we define a new quantity, the secondary
reproductive number, which predicts the epidemic threshold for highly
clustered and randomly wired scale-free networks, as well as for the Internet
graph as an example of a real-world scale-free network \cite{Faloutsos99}.

We consider the susceptible-infected-susceptible (SIS) model,
as a simple description of epidemic spreading in a population
\cite{Diekmann00}. Each individual in the population is either infected or
susceptible at any point in time. The potential infection pathways are
described by interpreting the individuals as the nodes of a network. The
time-discrete dynamics is defined by synchroneously updating the states of
all individuals with the following rules: If individual $A$ is infected at
time $t-1$, it is susceptible at time $t$. If, otherwise, individual $A$ is
susceptible at time $t-1$ and is connected to at least one individual
infected at the same time, then with probability $\lambda$ individual $A$ is
infected at time $t$. An important observable is the prevalence $\rho$. It is
the time average of the fraction of infected individuals reached after a
transient from the initial condition. Given a network, the only parameter of
the model is the infection probability $\lambda$. The information on the
global spreading of a disease is contained in the function $\rho(\lambda)$.

The individuals are connected by highly clustered scale-free networks
\cite{Klemm01}. They are constructed by iteratively adding nodes and links in
the following way: Generate a new node and connect it with all active nodes.
Set the new node active as well. Set inactive one of the active nodes. The
probability for deactivating node $i$ is inversely proportional to its
current number of links $k_i$. Close the iteration loop by generating the
next new node and so forth, until the network size reaches the desired value
$N$. Starting from an initial network of $m$ fully interconnected active
nodes, a network with an average degree $\langle k \rangle = 2m$ links per
node is generated. The degree distribution follows a power law $P(k)=2m^2
k^{-3}$, and the clustering coefficient $C=5/6$. Note that the deactivation
mechanism mentioned here is part only of the growth mechanism of the network.
It is not related to the dynamics of the SIS model which is applied after the
network has been constructed.

\begin{figure}
\centerline{\epsfig{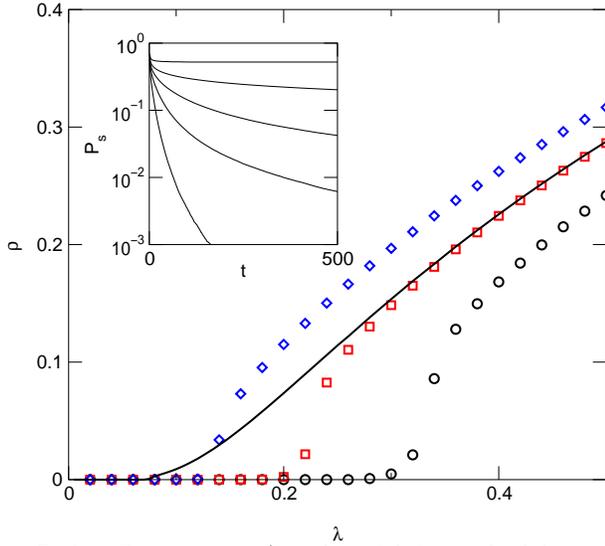}}
\caption{
Prevalence $\rho$ (fraction of infected individuals in the stationary
state) as a function of the spreading rate $\lambda$ for highly
clustered scale-free networks, with $\langle k \rangle = 4$ (circles),
6 (squares), and 10 (diamonds), and for random scale-free networks with
$\langle k \rangle = 6$ (solid curve). The
simulations have been run in networks containing $10^5$ nodes and
averaging over 100 different realizations. Inset: Survival probability, $P_s$, for a localized infection after $t$
time steps. Parameter values (from bottom to top) $\lambda=0.15$, 0.18, 0.2,
0.22, 0.25; and $\langle k \rangle = 6$. 
}
\label{fdis}
\end{figure}

By extensive simulations we have obtained the prevalence $\rho(\lambda)$  for
populations of $N=10^5$ individuals connected by highly clustered scale-free
networks. In Fig.~\ref{fdis} we plot the fraction of infected individuals in
the stationary state, $\rho$, for different values of the average
connectivity. Only when $\lambda$ is increased above a value $\lambda_c$
a significant prevalence is observed. The effect of the topological properties
of the highly clustered scale-free networks becomes clear when comparing
the shape of the prevalence curves with those obtained for randomly wired
scale-free networks. In the latter case no change of behavior is apparent
as the prevalence and its slope vary smoothly when $\lambda$ is increased.

Further insight into the behavior of epidemics in highly clustered scale-free
networks is gained from the time evolution of the survival probability $P_s$
shown in the inset of Fig.~\ref{fdis}. Taking initial conditions with exactly one
randomly chosen site infected, $P_s(t)$ is the fraction of realizations that
contain at least one infected site after $t$ time steps. For values of
$\lambda$ well below the threshold $\lambda_c$ the disease dies out
exponentially whereas for $\lambda$ above $\lambda_c$ the survival
probability $P_s$ approaches a non-zero plateau value.
The change of behavior from
rapid eradication to non-zero prevalence is observed at a finite value of the
transmission probability, independent of the system size. Thus the prevalence
of the SIS model in highly clustered scale-free networks undergoes a phase
transition at a finite critical value $\lambda_c$ of the transmission
probability. In other words, viruses with a low transmission probability do
not prevail in these networks.

In order to understand the role played by the
topology we consider the average connectivity of the
neighbors of a node $i$
\be
k_i^{nn} = \frac{1}{k_i} \sum_{j\in{\cal V}} k_j~,
\ee
where $k_j$ is the degree of node $j$ and the neighborhood of node
$i$ (\ie, the set of nodes directly connected to node $i$)
is called ${\cal V}$.

\begin{figure}
\centerline{\epsfig{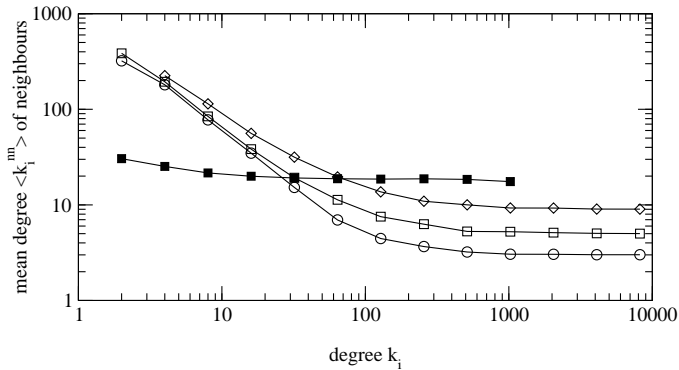}}
\caption{
Average degree of the neighbors of a node with connectivity $k$ in the
structured networks
with $\langle k \rangle =4$ (circles), 6 (open squares), 10 (diamonds). The
asymptotic values for large $k$ are $3.0 \pm 0.1$, $5.1 \pm 0.3$, and $9 \pm
1$ to be compared with the theoretical prediction $\langle
k_h^{nn}\rangle=\langle k \rangle-1 = 3$, 5, and 9 respectively (cf. Eq.~3). 
The filled squares is the average degree of the neighbors in random scale-free
networks with $\langle k \rangle =6$.
}
\label{fknn}
\end{figure}

The structure of the highly clustered scale-free networks gives rise to correlations between the
degree of a node and the degrees of its neighbors (see Fig.~\ref{fknn}). For
weakly connected nodes, $\langle k^{nn} \rangle$ decays. For
the hubs, $k \gg \langle k \rangle$, it reaches a constant value \cite{clust}
\be
\langle k_h^{nn} \rangle = \langle k \rangle-1~.
\label{hcsf}
\ee

\begin{figure}
\vskip 1cm
\centerline{\epsfig{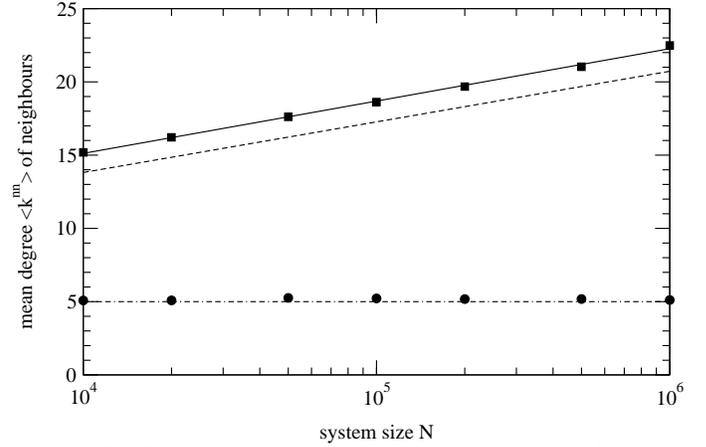}}
\caption{Dependence of the average degree of the neighbors of a node with
system size $N$. For the casse of highly clustered scale-free networks, the value has been obtained
averaging for nodes with $k > 1000$. The theoretical predictions ($\langle k
\rangle -1$) is
also plotted (dashed-dotted line).
For the case of the BA networks, the values are the average
over the full range of available connectivities.
The theoretical prediction $\langle k \rangle/4 \ln N$ is also plotted (dashed line). }
\label{fknn2}
\end{figure}

In order to unify the treatment of random and structured networks we also
calculate $\langle k^{nn}\rangle$ for random scale-free networks. If
$P_c(k^\prime|k)$ is the conditional probability that a link belonging to a
node with connectivity $k$ points to a node with connectivity $k^\prime$,
then
\be
\langle k^{nn} \rangle
= \sum_{k^\prime} k^\prime P_c(k^\prime |k)
= \sum_{k^\prime} \frac{(k^\prime)^2}{\langle k \rangle} P(k^\prime)
= \frac{\langle k^2 \rangle}{\langle k \rangle} ~,
\label{k2}
\ee
where we have used
$P_c(k^\prime |k) \propto k^\prime P(k^\prime)$ for random networks.
Now we specifically consider randomly wired scale-free networks
with the degree distribution $P(k) = 2m^2 k^{-3}$, the same distribution as
in the highly clustered scale-free networks considered before. The
networks are generated using the algorithm introduced in
Ref.~\cite{Barabasi99}. Ordering the nodes with respect to decreasing degree,
every node is identified by its rank $i$.  The degree of node $i$ is given
by 
\be
k_i(N) = \frac{\langle k \rangle}{2} \left(\frac{N}{i}\right)^{0.5}~.
\ee
Inserting 
$
\langle k^2 \rangle = \sum_{i=1}^{N} k_i^2 (N) = \langle k \rangle^2 /4
\ln N + {\cal O} (N^{-1})
$
into Eq.~(\ref{k2}) we obtain
\be
\langle k^{nn}\rangle = \frac{\langle k \rangle}{4} \ln N~,
\ee
independent, on average, of the node under consideration. This
independence is confirmed numerically, see Fig.~\ref{fknn}. It reflects
the absence of correlations in the connectivity. Figure \ref{fknn2}
shows the logarithmic dependence of $\langle k^{nn} \rangle$ on
system size, in contrast with the constant value obtained for the hubs
in the structured (highly clustered scale-free) networks.

Now the different connectivity of the hubs in the highly clustered and random
scale-free networks (both having the same degree distribution) is clear:
Whereas in the random case a hub is connected to other highly connected nodes,
in the highly clustered networks the hubs are almost exclusively connected to low
degree nodes. This difference will result essential for the epidemic dynamics.

But how is this topological property related to the transmission
threshold found of the SIS model?
Let us define the {\em secondary reproductive number} as
\be
R_2 = \lambda \langle k^{nn}_h \rangle~.
\label{R2}
\ee
We show below that the condition $R_2=1$ recovers a previous prediction for
the epidemic threshold in randomly wired networks
networks, and gives a good estimate for the highly clustered
scale-free networks and the Internet graph.

Previously, the behavior of the epidemics has been described in terms
of the {\em basic reproductive number}, $R_0$ \cite{May87}.
It is defined as the average number of
secondary infections produced by an infectious individual in a totally
susceptible population and indicates whether a disease can ever invade a
population. For random networks with broad degree distribution, the
basic reproductive number is given by
\be
R_0 = \lambda \frac{\langle k^2 \rangle}{\langle k\rangle}~.
\label{R0}
\ee
Only if $R_0$ is larger than unity the infection prevails.
Employing Eq.~(\ref{k2}) we find $R_0=R_2$, such that in randomly
wired networks the basic and secondary reproductive number coincide.
Therefore the condition $R_2=1$ recovers the standard prediction of
the epidemic threshold used in epidemiology, assuming random mixing
of the population.

For the highly clustered scale-free networks, applying the condition $R_2 = 1$  and using
Eq.~\ref{hcsf} predicts a threshold
\be
\lambda_c = \frac{1}{(\langle k \rangle -1)}~.
\ee
The onset of non-zero prevalence found numerically (Fig.\ \ref{fdis}) is in good
agreement with the prediction. Note that for the highly clustered scale-free networks
in general $R_2\neq R_0$. In particular, $R_0$ diverges with system size $N$ as
$\ln N$ leading to a false prediction of $\lambda_c=0$ in the limit of large
highly clustered scale-free networks.

\begin{figure}
\centerline{\epsfig{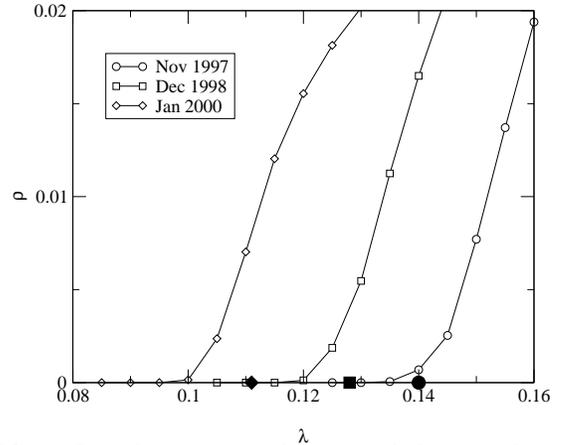}}
\caption{Prevalence $\rho$
as a function of the spreading rate $\lambda$ for the Internet graph at
three different times. The large filled symbols indicate the trasmission
threshold calculated according to the secondary reproductive number
(Eq.~(\ref{R2})). The value of $\langle k_h^{nn} \rangle$ has been
obtained as an average over the two largest hubs.}
\label{finf}
\end{figure}

In order to check the applicability of the secondary reproductive number to
empirical networks we investigate the Internet graph. We simulate the SIS model
in the network of the Autonomous Systems at three different time stages
of its evolution \cite{as}. Figure~\ref{finf} shows the prevalence of the SIS model as
a function of the transmission probability. The threshold values predicted by
the condition $1=\lambda_c \langle k_h^{nn} \rangle$ give a good estimate of
the minimum transmission rate above which the disease spreads. However, using
the basic reproductive number instead (Eq.~(\ref{R0})), gives threshold values
 0.012, 0.009, 0.007 for years 1997, 1998, 2000, respectively. This
understimates the threshold found in the simulations by at least one order of
magnitude. Similar to the highly clustered scale-free networks
the Internet graph displays considerable degree correlations \cite{Pastor01b}.
The mean connectivity in the neighborhoods of the hubs is much lower than
expected for random wiring. This explains the failure of the description by
the basic reproductive number which neglects the strong correlations. The
secondary reproductive number, however, gives a satisfactory prediction.

We have shown the existence of a finite epidemic
threshold in highly clustered scale-free networks in the limit of inifinite
system size. Our study has considered
for the first time scale-free networks with realistic topological properties
as a model for the potentially infective contacts between individuals or
nodes. We have conjectured that the value of the threshold is related to
the degree correlations in the network, such that the product of the
transmission probability $\lambda$ and the mean connectivity
$\langle k^{nn}_h\rangle$ of the neighbors of the hubs needs to exceed
unity for the epidemic to prevail. This criterion holds precisely for
highly clustered scale-free networks. For randomly wired scaele-free
networks it coincides with the standard prediction in epidemiology given
by the basic reproductive number. The transmission probability required for spreading on the real Internet graph is approximated well by our criterion,
whereas the basic reproductive number drastically underestimates the value.

The existence of an epidemic threshold in highly clustered scale-free
networks contrasts with the result for randomly wired networks, where
arbitrarily weak viruses show finite prevalence. This suggests that
the spreading of viruses in networks with scale-free degree distribution 
may be suppressed by non-random wiring. In particular, degree correlations
including the absence of direct connections between highly connected nodes,
may provide protection against epidemics.

\vspace*{2mm}

\begin{thebibliography}{99}
\bibitem{clust}
$\langle k_h^{nn} \rangle$
can be related to the local clustering $C(k)$ of a node of degree $k$:
$C(k) = (\langle k \rangle-2)/(k-1) + {\cal O}(k^{-2})$ \cite{Klemm02}.
For the hubs we have
$\langle k_h^{nn} \rangle \approx C(k) k + 1$,
where the first term accounts for the contribution of links to other nodes
in the neighborhood of the hub, and the last term reflects the link to the
hub itself. The limit $k \to \infty$ gives Eq.~(\ref{hcsf}).
\end{thebibliography}
\end{document}